\documentclass[12pt]{iopart}

\usepackage{iopams,graphicx}
\begin{document}

\title{Replica symmetry breaking in trajectory space for the trap model}

\author{Masahiko Ueda}
\address{Department of Basic Science, The University of Tokyo, Tokyo 153-8902, Japan}
\ead{ueda@complex.c.u-tokyo.ac.jp}
\author{Shin-ichi Sasa}
\address{Department of Physics, Kyoto University, Kyoto 606-8502, Japan}
\ead{sasa@scphys.kyoto-u.ac.jp}

\begin{abstract}
We study the localization in the one-dimensional trap model in terms of statistical mechanics of trajectories.
By numerically investigating overlap between trajectories of two particles on a common disordered potential, we find that there is a phase transition in the path ensemble.
We characterize the low temperature phase as a replica symmetry breaking phase in trajectory space.
\end{abstract}

%Uncomment for PACS numbers title message
% Keywords required only for MST, PB, PMB, PM, JOA, JOB? 
%\vspace{2pc}
%\noindent{\it Keywords}: Article preparation, IOP journals
% Uncomment for Submitted to journal title message
%\submitto{\JPA}
% Comment out if separate title page not required

\maketitle

\section{Introduction}
\label{sec:intro}
% subdiffusion
Subdiffusion is frequently observed in transport in non-equilibrium environments \cite{TMTet2004, WGRet2004, GolCox2006, SzyWei2009, BIKet2009, SenMar2010, WSTK2011, JTBet2011, PSCet2014}.
Theoretically, several mechanisms of subdiffusion have been proposed; continuous time random walk (CTRW), diffusion on fractal, and fractional Brownian motion \cite{BouGeo1990, MetKla2000, BJMB2011, HofFra2013}.
There are also many simple models which exhibit subdiffusion \cite{BouGeo1990}.
Inspired by the theoretical studies, concepts such as weak ergodicity breaking \cite{HBMB2008} have been used to analyze experimental data \cite{JTBet2011}.
The significance of subdiffusion in biological systems has also been discussed \cite{GolCox2006, BIKet2009}.

% trap model
One of the prominent models of subdiffusion is the trap model on a finite-dimensional lattice.
In this model, a quenched potential is defined on each site and a random walker is trapped by the potential during the waiting time.
A mean-field version of the trap model was originally introduced as an illustrative model of glassy behavior such as weak ergodicity breaking and aging \cite{Bou1992, MonBou1996}.
When the spatial dimension is larger than two, this model is effectively equivalent to CTRW, because each site is visited a finite number of times \cite{MonBou1996}.
In contrast, a one-dimensional version was shown to be qualitatively different from the mean-field version, since a random walker visits a given site many times \cite{BouGeo1990, BerBou2003}.
One of the interesting behaviors of the one-dimensional trap model is \textit{dynamical localization}, which means that there is a finite probability that independent particles are on the same site even after a very long waiting time.
This phenomenon is one simple example of localization in non-equilibrium states.

% RSB in trajectories
In our previous paper \cite{UedSas2015}, we developed a detection method of anomalous diffusion with localization in trajectory space with the concept of replica symmetry breaking (RSB) in trajectories.
RSB was proposed as a concept describing the low temperature phase of mean-field spin glass models \cite{MPV1987}, and means that two independent and identical systems with a common Hamiltonian have non-trivial overlap, which describes the similarity of the configurations of two systems.
We extended the concept of RSB to trajectory space, where the similarity of two independent and identical systems with a common dynamical rule is discussed.
Here, we apply this method to the trap model on a one-dimensional lattice, and show that the trap model exhibits RSB in trajectories.
That is, while we proposed a superdiffusive system as a model exhibiting RSB in trajectories in our previous study \cite{UedSas2015}, we claim that this phenomenon also occurs in a subdiffusive system such as the trap model.

% statistical mechanics of trajectories
From a more general perspective, we study anomalous diffusion in the trap model in terms of statistical mechanics of trajectories.
Statistical mechanics of trajectories was originally introduced in dynamical system theory \cite{BecSch1993}.
In this framework, cumulant generating function of a time-averaged quantity is regarded as the dynamical free energy, and the existence of a phase transition is discussed.
One of the successes of this framework is detection of the dynamical first-order phase transition characterized by the dynamical free energy of activity for glassy systems \cite{GJLPDW2007, HJGC2009, JacGar2010}.
Another application of this framework is calculation of the dynamical free energy of Lyapunov exponent \cite{GKLT2011, LLKT2013, LSTW2015}.
In this paper, we study the dynamical free energy of overlap \cite{UedSas2015}, which describes path-probability-measure concentration.

% organization of paper
This paper is organized as follows.
In section \ref{sec:model}, we introduce the model and review results of previous studies.
In section \ref{sec:eqstatmech}, we study equilibrium statistical mechanics of the trap model.
In section \ref{sec:trajectory}, we reformulate the problem of the trap model in terms of statistical mechanics of trajectories, by introducing overlap between trajectories of two systems.
In section \ref{sec:high}, we investigate the high temperature phase of the trap model.
In section \ref{sec:low}, we investigate the low temperature phase, by paying attention to the order of the large system-size limit and the large time limit.
Section \ref{sec:discussion} is devoted to concluding remarks.

\section{Model}
\label{sec:model}

% model and  quantities

We consider a single particle on a one-dimensional lattice
$\left\{ 1, \cdots, M \right\}$.
The position of the particle at time $t\in \left\{ 1, \cdots, \tau \right\}$
is denoted by $j_t$.
We impose the periodic boundary conditions for the lattice.
A quenched random variable $E_j>0$ is defined on each site $j$,
which is interpreted as an energy barrier. Once a particle has
escaped from the trap, it chooses one of the two neighboring
sites with the equal probability. The transition probability
of the particle is given by
\begin{eqnarray}
  T(j|j^\prime)
  &=& \left( 1 - e^{-\beta E_{j^\prime}} \right)
  \delta_{j,j^\prime} + \frac{1}{2} e^{-\beta E_{j^\prime}}
  \delta_{j,j^\prime+1} + \frac{1}{2} e^{-\beta E_{j^\prime}} \delta_{j,j^\prime-1},
\end{eqnarray}
where $\delta_{i,j}$ is the Kronecker delta,
and random variables $\left\{ E_j \right\}$
independently obey the exponential distribution
\begin{eqnarray}
 \rho(E) &=& e^{-E}.
\end{eqnarray}
The parameter $\beta$ represents the inverse temperature.
%where the Boltzmann constant is set to unity. 
The time evolution of the probability distribution
of the particle position is described by the Markov chain
\begin{eqnarray}
 P(j,t+1) &=& \sum_{j^\prime} T(j|j^\prime) P(j^\prime, t).
\end{eqnarray}
An initial distribution of the particle position is denoted by $P_0(j)$.
By noting the detailed balance condition, 
the stationary distribution is derived as 
\begin{eqnarray}
 P_\mathrm{eq}(j) &=& \frac{1}{Z_\mathrm{eq}} e^{\beta E_j},
\end{eqnarray}
where $Z_\mathrm{eq} \equiv \sum_{j=1}^M e^{\beta E_j}$
is the normalization constant.
We also define the probability distribution
of a particle trajectory $[j]\equiv \left( j_0, j_1, \cdots, j_\tau \right)$ by
\begin{eqnarray}
 \mathcal{P}[j] &=& P_0(j_0) \prod_{l=1}^\tau T\left( j_l|j_{l-1} \right).
 \label{eq:path_prob}
\end{eqnarray}
This model is a discrete-time version of the trap model.
Below we denote the expectations with respect to $\mathcal{P}[j]$ and $\rho(E)$
by $\left\langle \cdots \right\rangle$ and
$\mathbb{E}\left[ \cdots \right]$, respectively.

% basic properties

We briefly review basic properties of the trap model. 
First, the waiting time distribution $p(t)$ is proportional
to $t^{-(1+1/\beta)}$ in the large $t$ regime
\cite{BouGeo1990}. Since the average waiting time diverges
when $\beta>1$, the weak ergodicity breaking occurs in the low
temperature region $\beta>1$. 
Because of this, the mean-squared displacement of a particle
in the limit $M \to \infty$ changes its behavior at $\beta=1$ \cite{BouGeo1990}:
\begin{eqnarray}
 \mathbb{E}\left[ \left\langle \left( j_t-j_0 \right)^2 \right\rangle \right] \simeq \left\{
\begin{array}{ll}
 t &\quad (\beta<1) \\
 \frac{t}{\log(t)} &\quad (\beta=1) \\
 t^\frac{2}{1+\beta} &\quad (\beta>1).
\end{array}
\right.
\label{eq:trap_diffusion}
\end{eqnarray}
It should be noted that the subdiffusion is observed
in the low temperature region, and that its scaling is different from that in CTRW: $\mathbb{E}\left[ \left\langle \left( j_t-j_0 \right)^2 \right\rangle \right] \simeq t^{1/\beta}$.
Furthermore, the property in the low temperature
region becomes clearer by studying the participation ratio
\begin{eqnarray}
 Y_2(\tau) &\equiv& \sum_j P(j,\tau)^2.
\end{eqnarray}
This quantity in the limit $M \to \infty$ takes zero when particles
are not localized, while it takes a finite value when particles 
are localized in specific sites. The previous study showed that 
$Y_2$ takes a finite value  in the limits $M \rightarrow \infty$
and $\tau \rightarrow \infty$ when $\beta>1$  \cite{BerBou2003}.
It was called dynamical localization \cite{BerBou2003}.
Note that the two limits $M \rightarrow \infty$ and
$\tau \rightarrow \infty$ do not commute when $\beta>1$:
\begin{eqnarray}
 \lim_{\tau \rightarrow \infty} \lim_{M \rightarrow \infty} \mathbb{E}\left[ Y_2(\tau) \right] &\neq& \lim_{M \rightarrow \infty} \lim_{\tau \rightarrow \infty} \mathbb{E}\left[ Y_2(\tau) \right].
 \label{eq:Y_twolimit}
\end{eqnarray}
Because the right-hand side is equivalent to the participation ratio in the equilibrium state $\lim_{M \rightarrow \infty} \mathbb{E}\left[ \sum_j P_\mathrm{eq}(j)^2 \right]$, the result (\ref{eq:Y_twolimit}) implies that the system does not relax to the equilibrium state in the large $M$ limit.

\section{Equilibrium statistical mechanics}
\label{sec:eqstatmech}

% motivation

In this section, we study equilibrium statistical mechanics
of the trap model. 
We see that the equilibrium thermodynamic functions are ill-defined, which motivates for introducing statistical mechanics of trajectories.

% REM

First, we note that the equilibrium distribution
of the particle is equivalent to one of random energy models
\cite{BouMez1997} which are defined through some idealization
of a spin glass model. Explicitly, we consider a spin system
of $\tilde N$ sites.  Each spin configuration corresponds to
one site (particle position) of the lattice. That is, the number
of spin configurations $2^{\tilde N}$ is equal to $M$. By using 
the partition function of this system $Z_\mathrm{eq}$, the free
energy per spin is defined by
\begin{eqnarray}
  f &\equiv& - \frac{1}{\beta} \lim_{\tilde N\rightarrow \infty}
  \frac{1}{\tilde N} \mathbb{E}\left[ \log Z_\mathrm{eq} \right].
 \label{eq:f_REM}
\end{eqnarray}
It was found  \cite{BouMez1997} that this system exhibits
the first-order phase transition at $\beta=1$: 
\begin{eqnarray}
 f = \left\{
 \begin{array}{ll}
  -\frac{1}{\beta} \log 2 &\quad (\beta<1) \\
  -\log 2 &\quad (\beta>1).
 \end{array}
 \right.
\end{eqnarray}
This transition is characterized by the participation
ratio defined by
\begin{eqnarray}
 Y_2^\mathrm{(eq)} &\equiv& \sum_{j=1}^{2^{\tilde N}} P_\mathrm{eq}(j)^2.
\end{eqnarray}
It was shown that $Y_2^\mathrm{(eq)}$ is finite when $\beta>1$ \cite{BouMez1997}:
\begin{eqnarray}
 \lim_{{\tilde N}\rightarrow \infty} \mathbb{E}\left[ Y_2^\mathrm{(eq)} \right] = \left\{
 \begin{array}{ll}
  0 &\quad (\beta<1) \\
  1-\frac{1}{\beta} &\quad (\beta>1).
 \end{array}
 \right.
\end{eqnarray}
This result means that spin configurations for $\beta>1$ are frozen into
several stable configurations. 
These are correct and well-known results for a spin system.
However, the free energy (\ref{eq:f_REM}) does not describe thermodynamic behavior of the original particle model.
In the particle model, since $\tilde{N}$ is equal to $\log_2 M$ with volume $M$, the free energy is rewritten as
\begin{eqnarray}
  f &=& - \frac{1}{\beta} \lim_{M\rightarrow \infty} \frac{1}{\log_2 M} \mathbb{E}\left[ \log Z_\mathrm{eq} \right].
  \label{eq:f_REM_mod}
\end{eqnarray}
This quantity is not interpreted as the free energy of the particle per volume, but as the free energy per logarithm of volume, and therefore does not have clear meaning as a physical quantity for the particle model.

% equilibrium thermodynamics for particle model

In standard equilibrium thermodynamics, the thermodynamic limit of the particle system is defined as follows.
First, we consider identical $N$ particles in volume $M$.
We define the free energy of the $N$-particle system as $F_N=F_N(\beta,M,N)$.
Then, the thermodynamic limit for this system is defined by the free energy per volume with the density $\rho=N/M$ fixed:
\begin{eqnarray}
  f^\mathrm{(particle)} &=& \lim_{M\rightarrow \infty} \frac{1}{M} F_N.
\end{eqnarray}
This is the standard definition of thermodynamic limit of the free energy of the particle model, which is different from (\ref{eq:f_REM_mod}).

% idealizaed gas 

In order to consider the thermodynamics of the particle model according to the above prescription, we study non-interacting $N$ classical particles on a common disordered environment with large $N$.
The partition function of this system is given by
\begin{eqnarray}
 Z_N &=& g(N) \left( \sum_{j=1}^M e^{\beta E_j} \right)^N,
\end{eqnarray}
where the factor $g(N)$ depending only on $N$ is determined
so that the free energy is extensive.
For this system, we consider the thermodynamic limit
$N\rightarrow \infty$ and $M\rightarrow \infty$ with $N/M = \rho$ fixed.
The free energy of the system is defined as
\begin{eqnarray}
 F_N &\equiv& -\frac{1}{\beta} \mathbb{E}\left[ \log Z_N \right] \nonumber \\
 &=& -\frac{N}{\beta} \mathbb{E}\left[ \log Z_\mathrm{eq} \right] - \frac{1}{\beta} \log g(N).
\end{eqnarray}
For the high temperature phase $\beta<1$, $Z_\mathrm{eq}$
behaves as $Z_\mathrm{eq} \simeq \mathbb{E}\left[ Z_\mathrm{eq} \right] = M/(1-\beta)$ for large $M$,
and when we set $g(N)=1/N!$, we obtain an extensive $F_N$ as 
\begin{eqnarray}
  F_N &\simeq&
  -\frac{N}{\beta}
  \left[ \log\left( \frac{M}{N} \right) - \log(1-\beta) \right].
\end{eqnarray}
This result yields the internal energy using the relation
$U_N = \partial (\beta F_N)/\partial \beta$:
\begin{eqnarray}
 \lim_{M\rightarrow \infty} \frac{1}{M}U_N &=& - \rho \frac{1}{1-\beta},
 \label{eq:u_high}
\end{eqnarray}
which diverges at $\beta=1$.
The pressure is also obtained as
\begin{eqnarray}
 p &=& - \frac{\partial }{\partial M} F_N \nonumber \\
 &=& \frac{\rho}{\beta},
 \label{eq:p_high}
\end{eqnarray}
which is equivalent to the equation of state of the ideal gas.
Finally, the entropy density is calculated as
\begin{eqnarray}
 \lim_{M\rightarrow \infty} \frac{1}{M}S_N &=& \rho\log \frac{1}{\rho} - \rho\log (1-\beta) - \rho \frac{\beta}{1-\beta}.
 \label{eq:s_high}
\end{eqnarray}
It should be noted that the entropy becomes negative
at a temperature higher than $1/\beta=1$ and diverges to
$-\infty$ at $\beta=1$. The free energy also diverges at
$\beta=1$. Since the free energy is a continuous function of
the temperature, this singularity implies that there is
no thermodynamics for the low temperature phase $\beta>1$. 

Indeed, in the low temperature phase $\beta>1$,
the rescaled variable $\zeta \equiv Z_\mathrm{eq}/M^\beta$ has a limit distribution $P(\zeta)$ for large $M$,
because $z_j \equiv e^{\beta E_j}$ obeys
a power-law distribution \cite{BouGeo1990}.
In this case,
$F_N$ is written as 
\begin{eqnarray}
 F_N &\simeq& -N \log M -\frac{1}{\beta}\log g(N),
\end{eqnarray}
for which we cannot define the extensive free energy for any $g(N)$.
We also obtain the internal energy
\begin{eqnarray}
 \lim_{M\rightarrow \infty} \frac{1}{M}U_N &=& - \infty
 \label{eq:u_low}
\end{eqnarray}
and the pressure
\begin{eqnarray}
 p &=& \rho
 \label{eq:p_low}
\end{eqnarray}
for any $\beta >1$. That is, the low temperature phase cannot
be characterized by equilibrium statistical mechanics.

\section{Statistical mechanics of trajectories}
\label{sec:trajectory}

We study the low temperature phase by considering
statistical mechanics of trajectories. The most characteristic
quantity in this approach is 
the entropy rate associated with the path probability (\ref{eq:path_prob}), which is defined as
\begin{eqnarray}
 h_\mathrm{KS} &\equiv& -\lim_{\tau \rightarrow \infty} \frac{1}{\tau} \sum_{[j]} \mathcal{P}[j] \log \mathcal{P}[j].
\end{eqnarray}
This entropy, which represents the extent of variety of trajectories, has been referred to as the Kolmogorov-Sinai
entropy \cite{LAW2005, LAW2007}.
It was found that the entropy continuously becomes zero at $\beta=1$ in the limit $M \rightarrow \infty$ \cite{IwaSasup}:
\begin{eqnarray}
\fl \lim_{M \rightarrow \infty}h_\mathrm{KS} = \left\{
 \begin{array}{ll}
 (1-\beta)\left[ \log 2 + \beta - \int_0^\infty dE e^{-E} \left( e^{\beta E} - 1 \right) \log \left( 1-e^{-\beta E} \right) \right] &\quad (\beta<1)  \\
 0 &\quad (\beta>1).
 \end{array}
 \right.
 \label{eq:KS_cal}
\end{eqnarray}
This result means that the number of observed trajectories is less than any exponential function of the observation time in the low temperature phase and that the Kolmogorov-Sinai entropy is a continuous function of $\beta$.
Therefore, the singularity of the trap model at $\beta=1$ is identified with the second-order transition in the trajectory space.

% overlap

We analyze this transition at $\beta=1$ in terms of overlap between
two trajectories \cite{UedSas2015}. We prepare two systems with
a common realization of disorder $\left\{ E_j \right\}$.
We define overlap between the two trajectories $[j^{(1)}]$ and $[j^{(2)}]$ by
\begin{eqnarray}
 q &=& \frac{1}{\tau} \sum_{l=1}^\tau \delta_{j_l^{(1)}, j_l^{(2)}}.
\end{eqnarray}
This quantity describes how two trajectories are similar to each other,
and detects localization in the trajectory space.
In other words, when the system freezes into one or a few specific
trajectories, the expectation of overlap $\left\langle q \right\rangle$
becomes non-zero. Now, the statistical properties are described by the
generating function of the overlap $\left\langle e^{\tau \epsilon q} \right\rangle$.
This quantity is expressed as
\begin{eqnarray}
\fl \left\langle e^{\tau \epsilon q} \right\rangle &=& \sum_{[j^{(1)}], [j^{(2)}]} \mathcal{P}[j^{(1)}] \mathcal{P}[j^{(2)}] e^{\epsilon \sum_{l=1}^\tau \delta_{j_l^{(1)}, j_l^{(2)}}} \nonumber \\
\fl &=& \sum_{[j^{(1)}], [j^{(2)}]} P_0\left( j_0^{(1)} \right) P_0\left( j_0^{(2)} \right) \prod_{l=1}^\tau T_\epsilon \left( j_l^{(1)}, j_l^{(2)} | j_{l-1}^{(1)}, j_{l-1}^{(2)} \right),
 \label{eq:Psi_cal}
\end{eqnarray}
where we have introduced a transfer matrix
\begin{eqnarray}
 T_\epsilon \left( j_l^{(1)}, j_l^{(2)} | j_{l-1}^{(1)}, j_{l-1}^{(2)} \right) &\equiv& T\left( j_l^{(1)}|j_{l-1}^{(1)} \right) T\left( j_l^{(2)}|j_{l-1}^{(2)} \right) e^{\epsilon \delta_{j_l^{(1)}, j_l^{(2)}}}.
 \label{eq:TM}
\end{eqnarray}
Below we write $\textrm{\boldmath $j$} \equiv \left( j^{(1)}, j^{(2)} \right)$ collectively.
We also introduce the path ensemble of two independent systems biased by overlap as \cite{GJLPDW2007}
\begin{eqnarray}
 \mathcal{P}_\epsilon [\textrm{\boldmath $j$}] &\equiv& \frac{1}{\left\langle e^{\tau \epsilon q} \right\rangle} \mathcal{P}[j^{(1)}] \mathcal{P}[j^{(2)}] e^{\epsilon \sum_{l=1}^\tau \delta_{j_l^{(1)}, j_l^{(2)}}}
\end{eqnarray}
and the expectation of the quantity $A$ in the path ensemble $\mathcal{P}_\epsilon[\textrm{\boldmath $j$}]$ as
\begin{eqnarray}
 \left\langle A \right\rangle_\epsilon &\equiv& \sum_{[\textrm{\boldmath $j$}]} \mathcal{P}_\epsilon [\textrm{\boldmath $j$}] A = \frac{\left\langle A e^{\tau \epsilon q} \right\rangle}{\left\langle e^{\tau \epsilon q} \right\rangle}.
\end{eqnarray}
Particularly, the expectation of the overlap in the biased ensemble is given by
\begin{eqnarray}
 \left\langle q \right\rangle_\epsilon &=& \frac{\left\langle q e^{\tau \epsilon q} \right\rangle}{\left\langle e^{\tau \epsilon q} \right\rangle} = \frac{1}{\tau} \frac{\partial }{\partial \epsilon} \log\left\langle e^{\tau \epsilon q} \right\rangle.
 \label{eq:def_avq}
\end{eqnarray}
When $\epsilon$ is positive (negative), the trajectories of two independent systems with a positive (zero) overlap are more weighted in the biased ensemble.
When particles freeze into a few specific trajectories, this quantity experiences a discontinuous jump at $\epsilon=0$ in the limit $M \rightarrow \infty$ and $\tau \rightarrow \infty$:
\begin{eqnarray}
 \left\langle q \right\rangle_{+0} &\neq& \left\langle q \right\rangle_{-0}.
 \label{eq:def_RSB}
\end{eqnarray}
We identify this discontinuous behavior as the replica symmetry breaking
in trajectories. It should be noted that 
$\left\langle q \right\rangle_{+0} = \left\langle q \right\rangle_{-0}>0$
when particles freeze into one specific trajectory. In such a case, 
trajectories are localized but replica symmetry is not broken.

Before ending this section, we remark on the relation between our analysis and that of previous studies focusing on the participation ratio.
Because the trivial relation
\begin{eqnarray}
 \left\langle q \right\rangle = \frac{1}{\tau} \sum_{l=1}^\tau Y_2(l)
 \label{eq:q-Y2}
\end{eqnarray}
holds, analysis of the average overlap is formally equivalent to analysis of the participation ratio.
However, we are interested in the generating function of overlap $\left\langle e^{\tau \epsilon q} \right\rangle$, which includes more information than the first moment $\left\langle q \right\rangle$.
Although the participation ratio $Y_2$ detect localization, it cannot distinguish whether localized states are unique or not.
In contrast, when we use the generating function of overlap $\left\langle e^{\tau \epsilon q} \right\rangle$, we can distinguish it through discontinuity of the first derivative.
This is the main advantage of our analysis based on RSB.

In the following sections, we investigate properties of the cumulant
generating function $\lim_{M, \tau \rightarrow \infty} (1/\tau) \log\left\langle e^{\tau \epsilon q} \right\rangle$ or
its derivative (\ref{eq:def_avq}) in the high temperature phase
and the low temperature phase.

\section{High temperature phase}
\label{sec:high}
In this section, we study the high temperature phase $\beta<1$.
The generating function $\left\langle e^{\tau \epsilon q} \right\rangle$ can be calculated by using the largest eigenvalue of the transfer matrix.
For finite $M$, we define the scaled cumulant generating function of overlap by
\begin{eqnarray}
 \psi(\epsilon) &\equiv& \lim_{\tau \rightarrow \infty} \frac{1}{\tau} \log \left\langle e^{\tau \epsilon q} \right\rangle,
 \label{eq:cum}
\end{eqnarray}
which is also referred to as the dynamical free energy.
By using the largest eigenvalue $\lambda(\epsilon)$ of the transfer matrix $T_\epsilon(\textrm{\boldmath $j$}|\textrm{\boldmath $j^\prime$})$ given by (\ref{eq:TM}), $\psi(\epsilon)$ is expressed as
\begin{eqnarray}
 \psi(\epsilon) &=& \log \lambda(\epsilon).
 \label{eq:cum-eig}
\end{eqnarray}
The eigenvalue equation is described as
\begin{eqnarray}
 \sum_{\textrm{\boldmath $j^\prime$}} T_\epsilon \left( \textrm{\boldmath $j$} | \textrm{\boldmath $j^\prime$} \right) \Phi_\epsilon(\textrm{\boldmath $j^\prime$}) &=& \lambda(\epsilon) \Phi_\epsilon(\textrm{\boldmath $j$}),
 \label{eq:eig_eq}
\end{eqnarray}
where $\Phi_\epsilon(\textrm{\boldmath $j$})$ is the eigenfunction corresponding to the eigenvalue $\lambda(\epsilon)$.
When $\epsilon=0$, we obtain $\lambda(0)=1$ and $\Phi_0(\textrm{\boldmath $j$})=P_\mathrm{eq}(\textrm{\boldmath $j$}) \equiv P_\mathrm{eq}(j^{(1)})P_\mathrm{eq}(j^{(2)})$.
Now we study the first moment $\left\langle q \right\rangle_0$.
By expanding the eigenvalue equation (\ref{eq:eig_eq}) in $\epsilon$ around $\epsilon=0$ and collecting terms proportional to $\epsilon$, we obtain
\begin{eqnarray}
\fl \sum_\textrm{\boldmath $j^\prime$} \left. \frac{\partial T_\epsilon}{\partial \epsilon} \left( \textrm{\boldmath $j$} | \textrm{\boldmath $j^\prime$} \right) \right|_{\epsilon=0} P_\mathrm{eq}(\textrm{\boldmath $j^\prime$}) + \sum_\textrm{\boldmath $j^\prime$} T\left( \textrm{\boldmath $j$} | \textrm{\boldmath $j^\prime$} \right) \left. \frac{\partial \Phi_\epsilon}{\partial \epsilon} (\textrm{\boldmath $j^\prime$})\right|_{\epsilon=0} &=& \left. \frac{\partial \lambda}{\partial \epsilon}\right|_{\epsilon=0} P_\mathrm{eq}(\textrm{\boldmath $j$}) + \left. \frac{\partial \Phi_\epsilon}{\partial \epsilon}(\textrm{\boldmath $j$}) \right|_{\epsilon=0}.
 \label{eq:TM_first}
\end{eqnarray}
By calculating the sum $\sum_\textrm{\boldmath $j$}$ of both sides, we obtain
\begin{eqnarray}
 \left. \frac{\partial \lambda}{\partial \epsilon}\right|_{\epsilon=0} &=& \sum_{\textrm{\boldmath $j$}} \sum_{\textrm{\boldmath $j^\prime$}} \left. \frac{\partial T_\epsilon}{\partial \epsilon} \left( \textrm{\boldmath $j$} | \textrm{\boldmath $j^\prime$} \right) \right|_{\epsilon=0} P_\mathrm{eq}(j^{\prime(1)})P_\mathrm{eq}(j^{\prime(2)}) \nonumber \\
 &=& \sum_j P_\mathrm{eq}(j)^2.
 \label{eq:eig_1}
\end{eqnarray}
This is the participation ratio in the equilibrium state:
\begin{eqnarray}
 Y_2^\mathrm{(eq)} &\equiv& \sum_j P_\mathrm{eq}(j)^2.
 \label{eq:def_Yeq}
\end{eqnarray}
We note that the first moment $\left\langle q \right\rangle_0$ is described as
\begin{eqnarray}
 \left\langle q \right\rangle_0 &=& \left. \frac{\partial \psi}{\partial \epsilon}\right|_{\epsilon=0} = \left. \frac{\partial \lambda}{\partial \epsilon}\right|_{\epsilon=0} = Y_2^\mathrm{(eq)}.
\end{eqnarray}
Therefore, the average overlap at $\epsilon=0$ is given by the participation ratio in the equilibrium state.
%This result indicates that previous studies focusing on the participation ratio are equivalent to analysis only dealing with the average overlap, while we are interested in the cumulant generating function of overlap $\left\langle e^{\tau \epsilon q} \right\rangle$

Now, because the configurational average of $Y_2^\mathrm{(eq)}$ was calculated exactly \cite{Der1997, BouMez1997}, we obtain
\begin{eqnarray}
 \lim_{M \rightarrow \infty} \mathbb{E}\left[ \left\langle q \right\rangle_0 \right] &=& 0
 \label{eq:ov_eq_high}
\end{eqnarray}
for $\beta<1$.
Therefore, there is no localization in trajectory space.
This result implies that RSB in trajectories does not occur in the high temperature phase $\beta<1$.

\section{Low temperature phase}
\label{sec:low}
In this section, we study the low temperature phase $\beta>1$.
Because of the trivial relation (\ref{eq:q-Y2}) and the previous result (\ref{eq:Y_twolimit}), we find that the two limits $M \rightarrow \infty$ and $\tau \rightarrow \infty$ do not commute in the calculation of $\left\langle q \right\rangle$.
We conjecture that this non-commutativity appears even in the calculation of the cumulant generating function $\lim_{M, \tau \rightarrow \infty} (1/\tau) \log \left\langle e^{\tau \epsilon q} \right\rangle$.
We investigate two cases separately.

\subsection{$M \rightarrow \infty$ after $\tau \rightarrow \infty$}
\label{subsec:mat}
We first study the cumulant generating function $\lim_{M \rightarrow \infty}\lim_{\tau \rightarrow \infty} (1/\tau) \log\left\langle e^{\tau \epsilon q} \right\rangle$.
This quantity can be calculated by the same method used in section \ref{sec:high}.
The cumulant generating function for finite $M$ is expressed by using the largest eigenvalue of the transfer matrix as (\ref{eq:cum-eig}).

First, we focus on the first moment $\left\langle q \right\rangle_0$.
This quantity is expressed as $\left\langle q \right\rangle_0 = Y_2^\mathrm{(eq)}$ as in section \ref{sec:high}.
The configurational average of $Y_2^\mathrm{(eq)}$ was calculated exactly \cite{Der1997, BouMez1997}, and we obtain
\begin{eqnarray}
 \lim_{M \rightarrow \infty} \mathbb{E}\left[ \left\langle q \right\rangle_0 \right] &=& 1-\frac{1}{\beta}
 \label{eq:ov_eq}
\end{eqnarray}
for $\beta>1$.
Therefore, in the low temperature phase $\beta>1$, localization in trajectory space occurs.

\begin{figure}[t]
\includegraphics[clip, width=8.0cm]{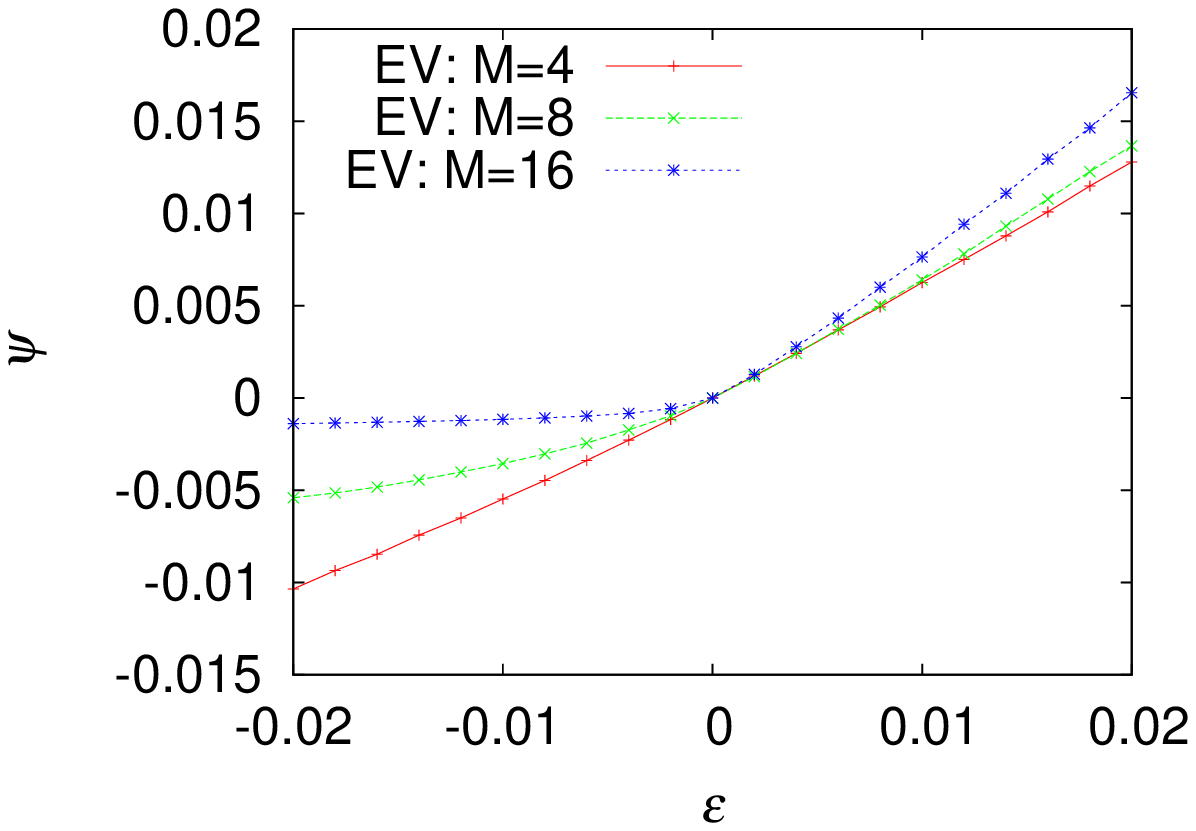}
\includegraphics[clip, width=8.0cm]{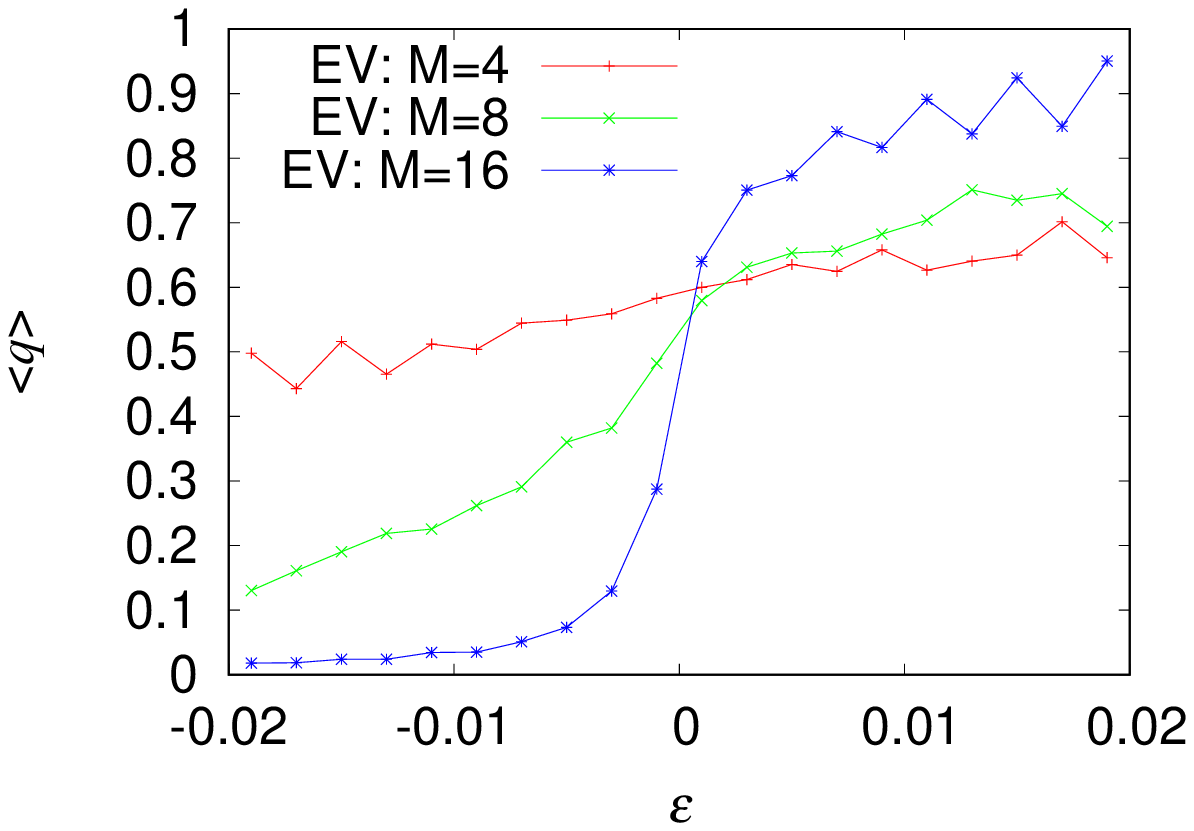}
\caption{The cumulant generating function $\psi(\epsilon)$ (left) and its numerical differentiation $\left\langle q \right\rangle_\epsilon$ (right) at $\beta=2$ for various system sizes.}
\label{fig:trap_tmatrix_eig_0.500000}
\end{figure}

In order to investigate whether the RSB occurs or not in the low temperature phase, we numerically estimate $\left\langle q \right\rangle_\epsilon$ from $\psi(\epsilon)$ which is obtained as a solution of the eigenvalue equation (\ref{eq:eig_eq}).
As an example, we consider the case $\beta=2$, and we prepared $N_\mathrm{d}$ realizations of the random potential, where $N_\mathrm{d}=10000$.
The cumulant generating function $\psi(\epsilon)$ is displayed in the left side of Fig. \ref{fig:trap_tmatrix_eig_0.500000}.
We observe that it becomes more bent at $\epsilon=0$ as system size $M$ increases.
Because $\left\langle q \right\rangle_\epsilon$ corresponds to the first derivative of $\psi(\epsilon)$, this result suggests the discontinuous behavior of $\left\langle q \right\rangle_\epsilon$ at $\epsilon=0$ in the limit $M \rightarrow \infty$.
This claim is confirmed more explicitly by directly calculating the numerical differentiation of the generating function $\psi(\epsilon)$, which is displayed on the right side of Fig. \ref{fig:trap_tmatrix_eig_0.500000}.
That is, we conjecture that RSB occurs in the sense (\ref{eq:def_RSB}).

\subsection{$M \rightarrow \infty$ before $\tau \rightarrow \infty$}
\label{subsec:mbt}
We study the cumulant generating function $\lim_{\tau \rightarrow \infty}\lim_{M \rightarrow \infty} (1/\tau) \log\left\langle e^{\tau \epsilon q} \right\rangle$.
It has been known that there exists a non-trivial limit of the participation ratio $\lim_{\tau \rightarrow \infty} \lim_{M \rightarrow \infty} \mathbb{E}\left[ Y_2(\tau) \right] \simeq 2(1-1/\beta)/3$ \cite{BerBou2003}.
In numerical simulation, this limit value is estimated for the system with finite $M$ and $\tau$ by analyzing the time region $\tau \ll \tau_\mathrm{eq}(M)$ with an equilibration time $\tau_\mathrm{eq}(M)$.
Here, $\tau_\mathrm{eq}(M)$ is estimated from the subdiffusion (\ref{eq:trap_diffusion}), that is, the time needed for a particle to diffuse a distance $M$:
\begin{eqnarray}
 \tau_\mathrm{eq}(M) &\sim& M^{1+\beta}.
 \label{eq:eqtime_low}
\end{eqnarray}
The existence of the non-trivial limit for $\mathbb{E}\left[ Y_2(\tau) \right]$ means that the system never relaxes to the equilibrium state with the participation ratio $\mathbb{E}\left[ Y_2^\mathrm{(eq)} \right]$ in the large $M$ limit.
Here, we investigate the generating function $(1/\tau) \log\left\langle e^{\tau \epsilon q} \right\rangle$ in the time region $\tau \ll \tau_\mathrm{eq}(M)$.

First, we calculate the average overlap in the biased ensemble $\left\langle q \right\rangle_\epsilon$.
This quantity is computed by the simple iterative calculation of a transfer matrix.
We introduce the quantity
\begin{eqnarray}
 Z_\epsilon(\textrm{\boldmath $j$},\tau) &\equiv& \left\langle \delta_{\textrm{\boldmath $j$}, \textrm{\boldmath $j$}_\tau} e^{\tau \epsilon q} \right\rangle.
\end{eqnarray}
Then, $\left\langle q \right\rangle_\epsilon$ is expressed by using $Z_\epsilon(\textrm{\boldmath $j$},\tau)$ as
\begin{eqnarray}
 \left\langle q \right\rangle_\epsilon &=& \frac{1}{\tau} \frac{\sum_{\textrm{\boldmath $j$}} \frac{\partial }{\partial \epsilon} Z_\epsilon(\textrm{\boldmath $j$},\tau)}{\sum_{\textrm{\boldmath $j$}} Z_\epsilon(\textrm{\boldmath $j$},\tau)}.
\end{eqnarray}
The quantities $Z_\epsilon(\textrm{\boldmath $j$},\tau)$ and $\partial Z_\epsilon(\textrm{\boldmath $j$},\tau)/\partial \epsilon$ satisfy the recurrence relation
\begin{eqnarray}
\fl Z_\epsilon(\textrm{\boldmath $j$},\tau) &=& \sum_{\textrm{\boldmath $j$}^\prime} T_\epsilon\left( \textrm{\boldmath $j$} | \textrm{\boldmath $j^\prime$} \right) Z_\epsilon(\textrm{\boldmath $j$}^\prime,\tau-1), \\
\fl \frac{\partial }{\partial \epsilon}Z_\epsilon(\textrm{\boldmath $j$},\tau) &=& \sum_{\textrm{\boldmath $j$}^\prime} \delta_{j^{(1)}, j^{(2)}} T_\epsilon\left( \textrm{\boldmath $j$} | \textrm{\boldmath $j^\prime$} \right) Z_\epsilon(\textrm{\boldmath $j$}^\prime,\tau-1) + \sum_{\textrm{\boldmath $j$}^\prime} T_\epsilon\left( \textrm{\boldmath $j$} | \textrm{\boldmath $j^\prime$} \right) \frac{\partial }{\partial \epsilon} Z_\epsilon(\textrm{\boldmath $j$}^\prime,\tau-1),
\end{eqnarray}
with the initial condition $Z_\epsilon(\textrm{\boldmath $j$},0)=P_0\left( j^{(1)} \right)P_0\left( j^{(2)} \right)$ and $\partial Z_\epsilon(\textrm{\boldmath $j$},0)/\partial \epsilon = 0$.
Therefore, by calculating these two quantities iteratively, we can directly obtain $\left\langle q \right\rangle_\epsilon$.

\begin{figure}[t]
\includegraphics[clip, width=8.0cm]{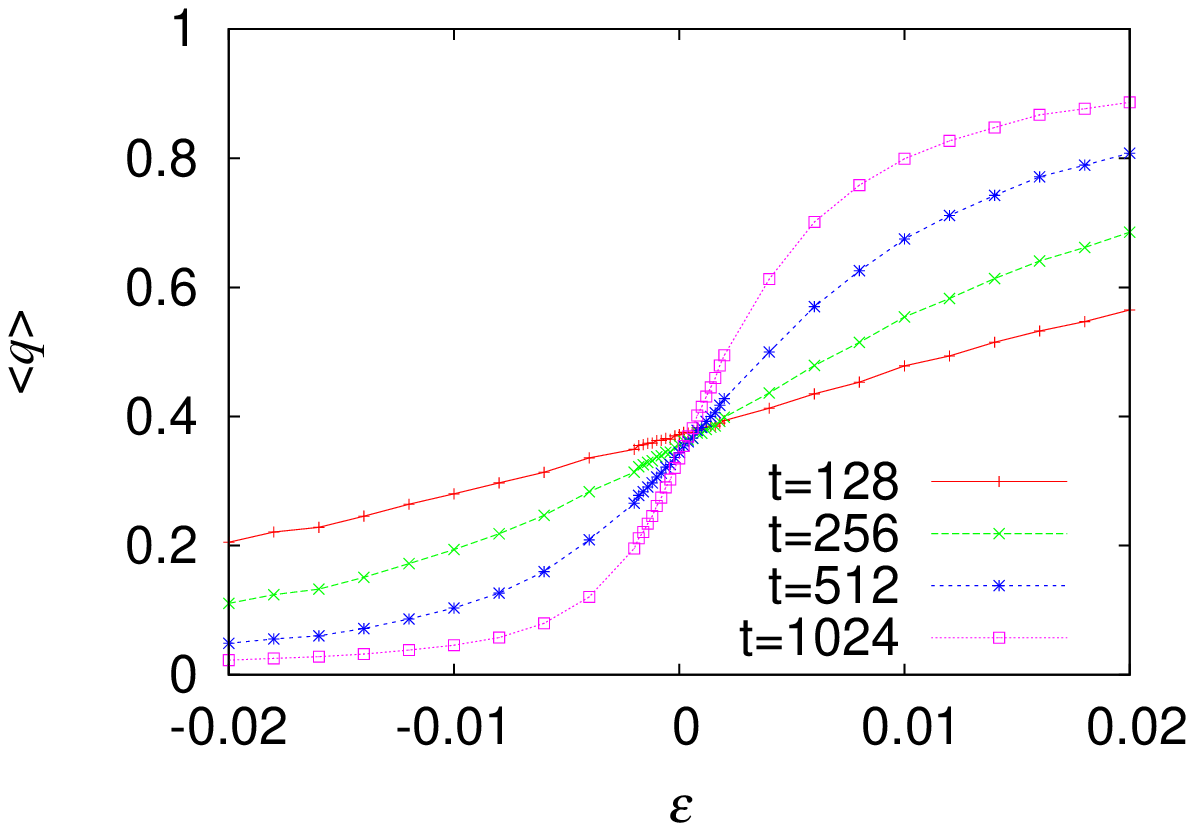}
\includegraphics[clip, width=8.0cm]{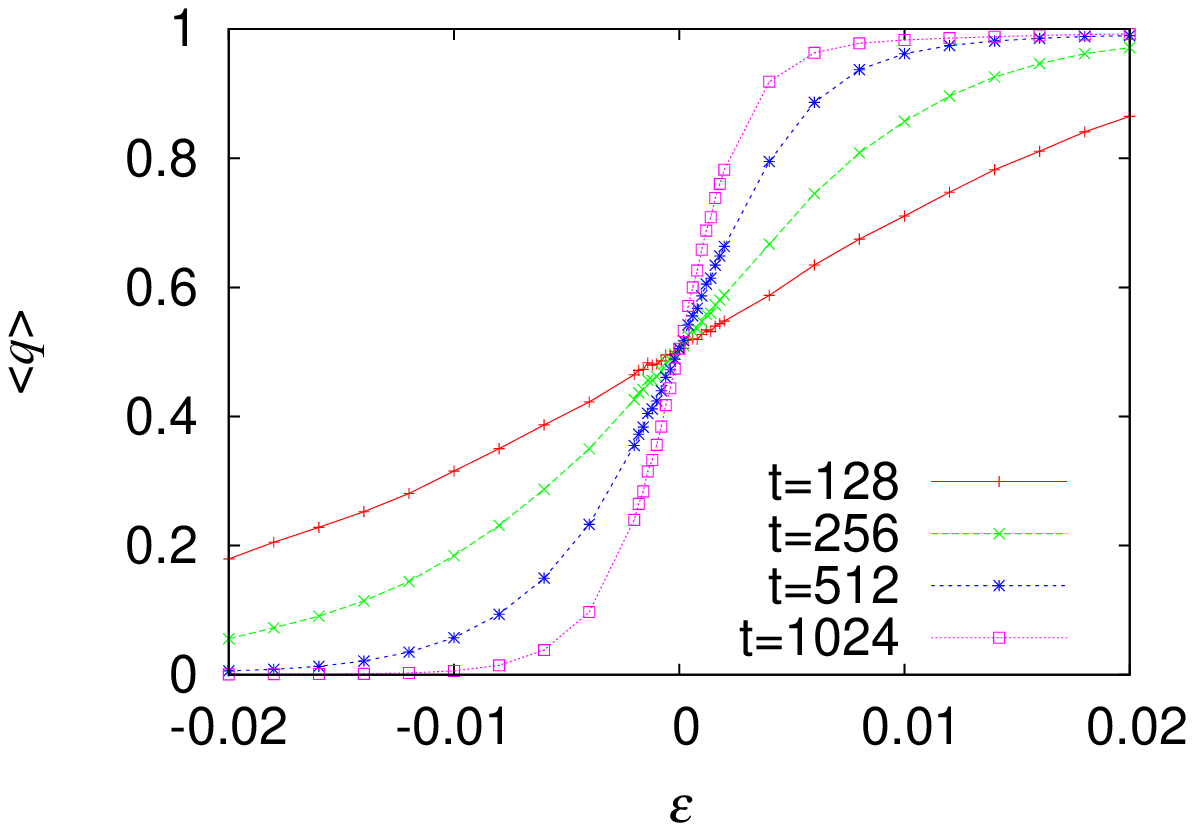}
\caption{The average overlap in the biased ensemble $\left\langle q \right\rangle_\epsilon$ for $M=64$ and $\beta=2$ under the initial distributions $P_0(j)=\delta_{j,1}$ (left) and $P_\mathrm{eq}(j)$ (right).}
\label{fig:ovtmatrix_trapv5_0.500000_64}
\end{figure}

The results for the initial distributions $P_0(j)=\delta_{j,1}$ and $P_0(j)=P_\mathrm{eq}(j)$ are displayed in Fig. \ref{fig:ovtmatrix_trapv5_0.500000_64}.
In this calculation, we fix $\beta=2$ and $N_\mathrm{d}=10000$, which is the same as that for Fig. \ref{fig:trap_tmatrix_eig_0.500000}.
\begin{figure}[t]
\includegraphics[clip, width=8.0cm]{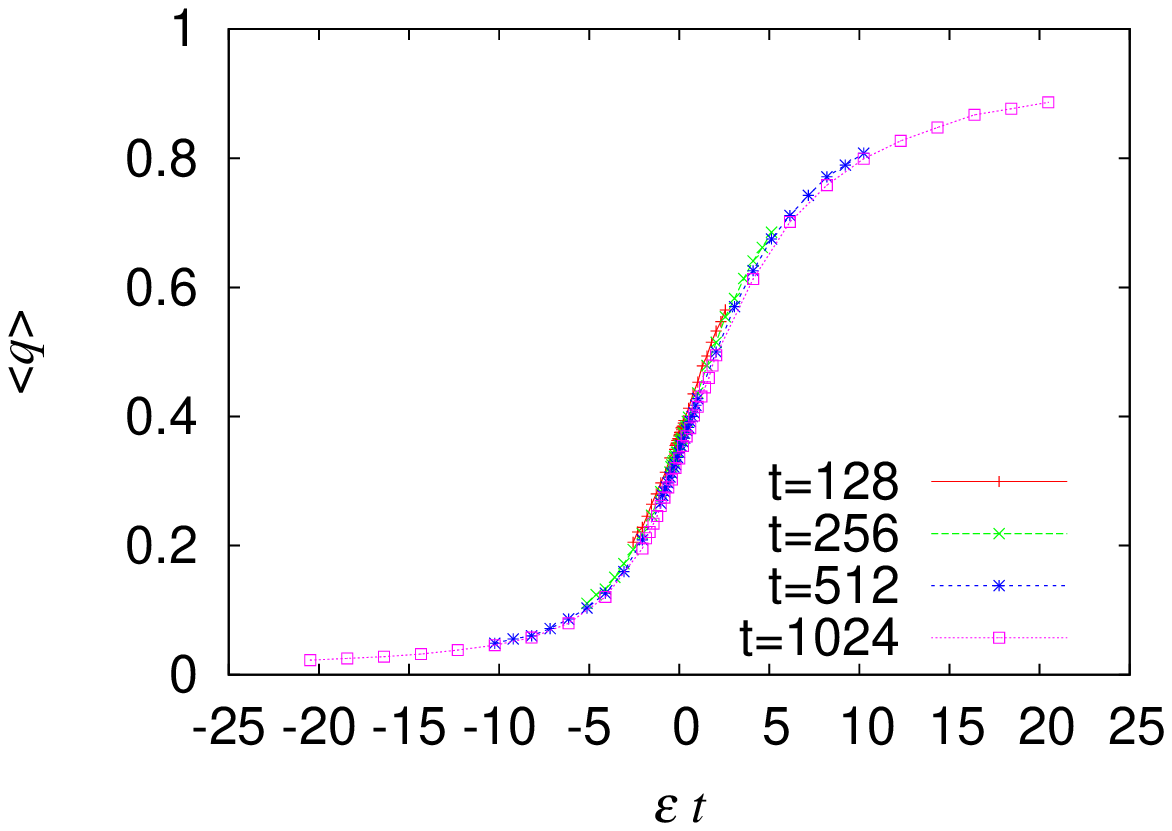}
\includegraphics[clip, width=8.0cm]{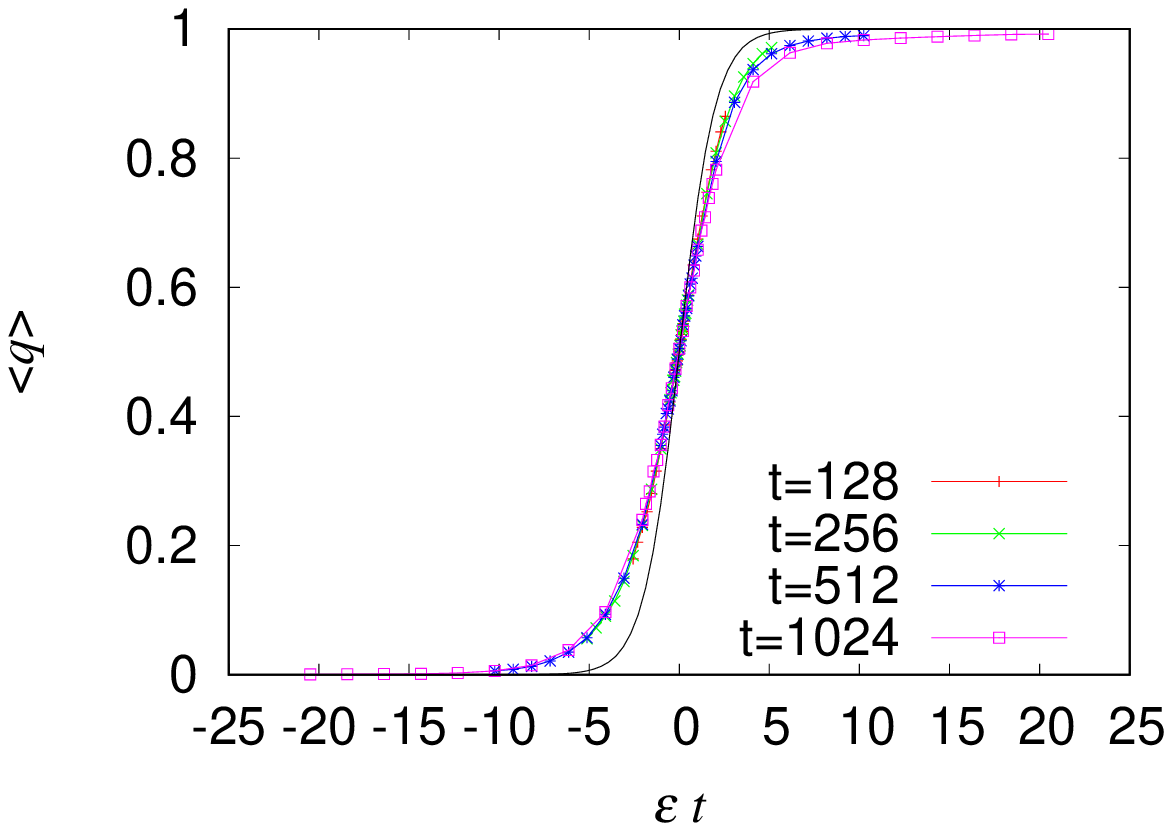}
\caption{Plot of the average overlap in the biased ensemble $\left\langle q \right\rangle_\epsilon$ versus $\epsilon t$ for $M=64$ and $\beta=2$ under the initial distributions $P_0(j)=\delta_{j,1}$ (left) and $P_\mathrm{eq}(j)$ (right). The solid curve on the right side corresponds to (\ref{eq:f_ana}).}
\label{fig:ovtmatrix_trapv5_0.500000_64_scale}
\end{figure}
Recalling the equilibration time $\tau_\mathrm{eq}(M)$ given in (\ref{eq:eqtime_low}), we focus on the time region $t \ll \tau_\mathrm{eq}(M)$ with $M=64$.
We observe that the slope at $\epsilon=0$ becomes larger and larger as time $t$ increases.
Particularly, when the horizontal axis is rescaled as $\epsilon t$ as in Fig. \ref{fig:ovtmatrix_trapv5_0.500000_64_scale}, the data collapse well.
These results show that $\lim_{M \rightarrow \infty} \left\langle q \right\rangle_\epsilon$ is expressed as a scaling form $\lim_{M \rightarrow \infty} \left\langle q \right\rangle_\epsilon = f(\epsilon t)$.
This implies $\lim_{\epsilon \rightarrow +0} \lim_{t \rightarrow \infty} \lim_{M \rightarrow \infty} \left\langle q \right\rangle_\epsilon = f(+\infty)$, while $\lim_{\epsilon \rightarrow -0} \lim_{t \rightarrow \infty} \lim_{M \rightarrow \infty} \left\langle q \right\rangle_\epsilon = f(-\infty)$.
Because $f(+\infty)\neq f(-\infty)$, we conclude that $\lim_{t \rightarrow \infty} \lim_{M \rightarrow \infty} \left\langle q \right\rangle_\epsilon$ is discontinuous at $\epsilon=0$.
Therefore, RSB in trajectories (\ref{eq:def_RSB}) occurs.
It should be noted that behavior of the scaling function depends on the initial condition, reflecting the fact that the system never relaxes to the equilibrium state in the large $M$ limit.
We also remark that, in the case of initial condition $P_0(j)=P_\mathrm{eq}(j)$, the expected form of $f(\epsilon t)$ is 
\begin{eqnarray}
 f(\epsilon t) = \frac{(\beta-1) e^{\epsilon t}}{(\beta-1)e^{\epsilon t} + 1}, 
 \label{eq:f_ana}
\end{eqnarray}
since particles freeze into the initial positions and the generating function is approximated as
\begin{eqnarray}
 \left\langle e^{t \epsilon q} \right\rangle &\simeq& \sum_{j_0^{(1)}, j_0^{(2)}} P_\mathrm{eq}\left( j_0^{(1)} \right) P_\mathrm{eq}\left( j_0^{(2)} \right) e^{\epsilon t \delta_{j_0^{(1)}, j_0^{(2)}}} \nonumber \\
 &=& 1 - Y_2^\mathrm{(eq)} + Y_2^\mathrm{(eq)} e^{\epsilon t}
\end{eqnarray}
in the large-$M$ limit.
Our numerical data deviate from this expression as shown in Fig. \ref{fig:ovtmatrix_trapv5_0.500000_64_scale}.
This may come from the finite-size effect of $M$.

We remark on the scaling form $\lim_{M \rightarrow \infty} \left\langle q \right\rangle_\epsilon = f(\epsilon t)$.
When we regard a trajectory of a particle as a directed polymer, such a scaling relation corresponds to that of RSB \cite{Mez1990}.
A directed polymer is a one-dimensional object, and its transverse and longitudinal directions are labeled by $x$ and $t$, respectively \cite{HalZha1995}.
Its Hamiltonian is described by a functional $H[x]$, and we focus on equilibrium statistical mechanics of polymers.
Let us consider the Hamiltonian of two polymers coupled by an interaction $t\epsilon q$: $H_\epsilon[\textrm{\boldmath $x$}] \equiv H\left[ x^{(1)} \right]+H\left[ x^{(2)} \right]+t\epsilon q$.
Generally, the effect of the interaction between two copies in the Hamiltonian increases as $\epsilon t$.
When the fluctuation of the one-polymer free energy among metastable states increases as $t^\omega$, the order of $\epsilon$ is estimated from the relation $\epsilon t \sim t^{\omega}$, and the scaling form $\lim_{M \rightarrow \infty} \left\langle q \right\rangle_\epsilon = f(\epsilon t^{1-\omega})$ is expected to be observed.
The $(1+1)$-dimensional directed polymers in a Gaussian random potential correspond to the case of $\omega=1/3$, and the scaling form $\lim_{M \rightarrow \infty} \left\langle q \right\rangle_\epsilon = f(\epsilon t^{2/3})$ was indeed observed \cite{Mez1990}.
In contrast, our case corresponds to $\omega=0$, which means that the free energy fluctuation does not increase with $t$ and several metastable states with the same free energy value coexist, that is, replica symmetry breaking.
Therefore, our numerical result surely suggests RSB in trajectories.

\section{Concluding remarks}
\label{sec:discussion}
\begin{figure}[t]
\includegraphics[clip, width=8.0cm]{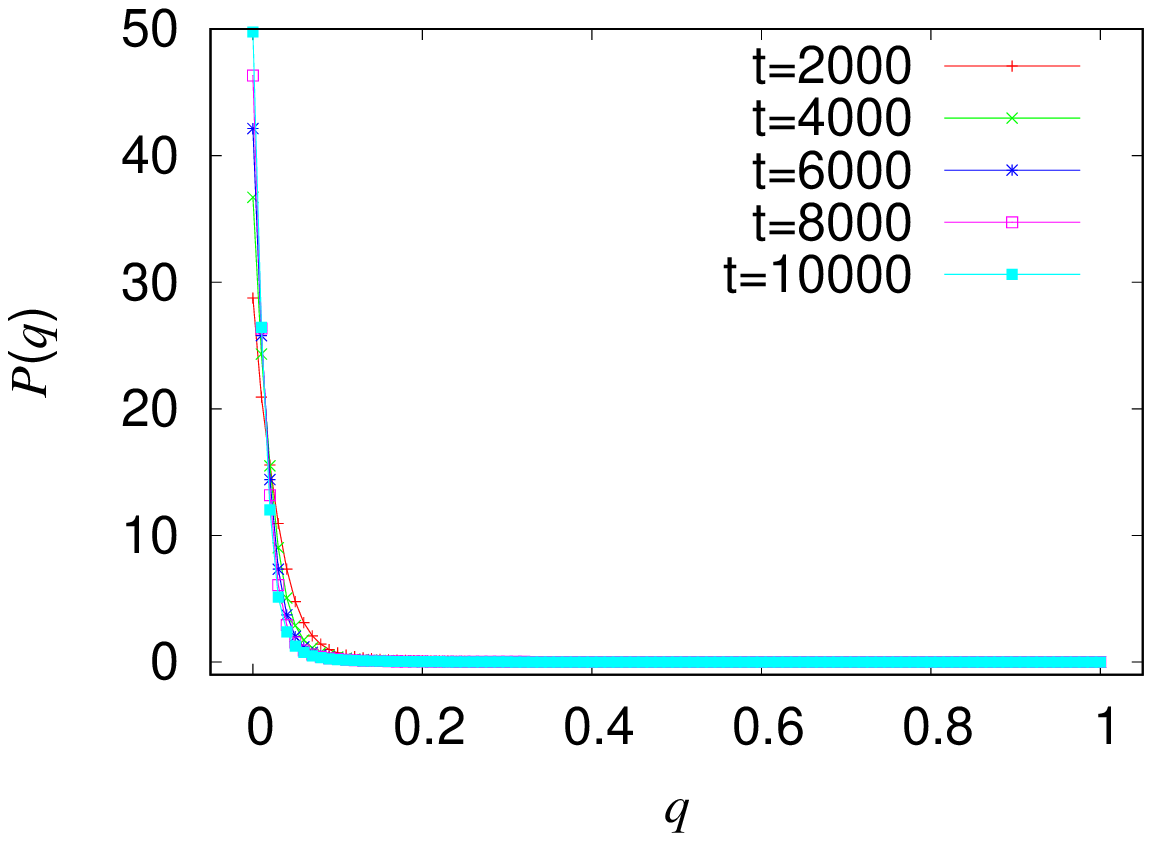}
\includegraphics[clip, width=8.0cm]{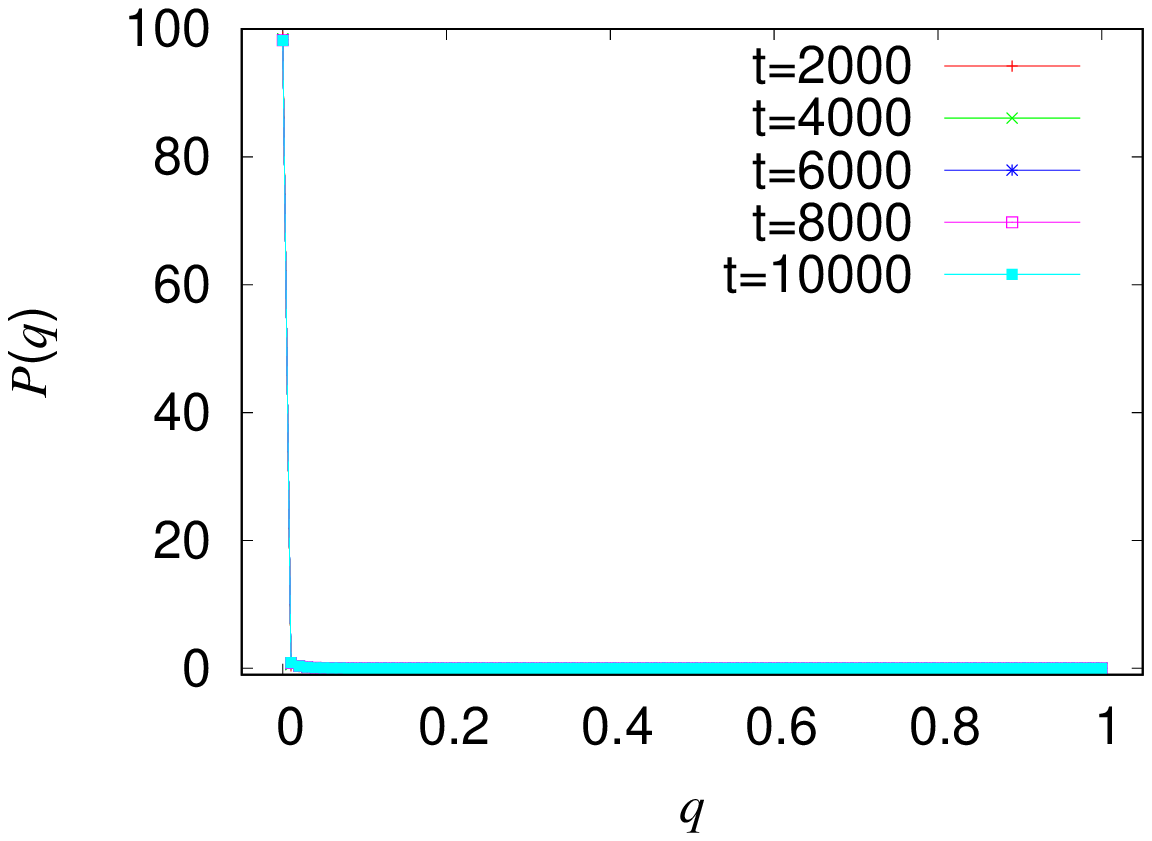}
\caption{The overlap distribution $P(q)$ at $\beta=0.5$ under the initial distributions $P_0(j)=\delta_{j,1}$ (left) and $P_\mathrm{eq}(j)$ (right).}
\label{fig:ovhisttrap_2.0}
\end{figure}

\begin{figure}[t]
\includegraphics[clip, width=8.0cm]{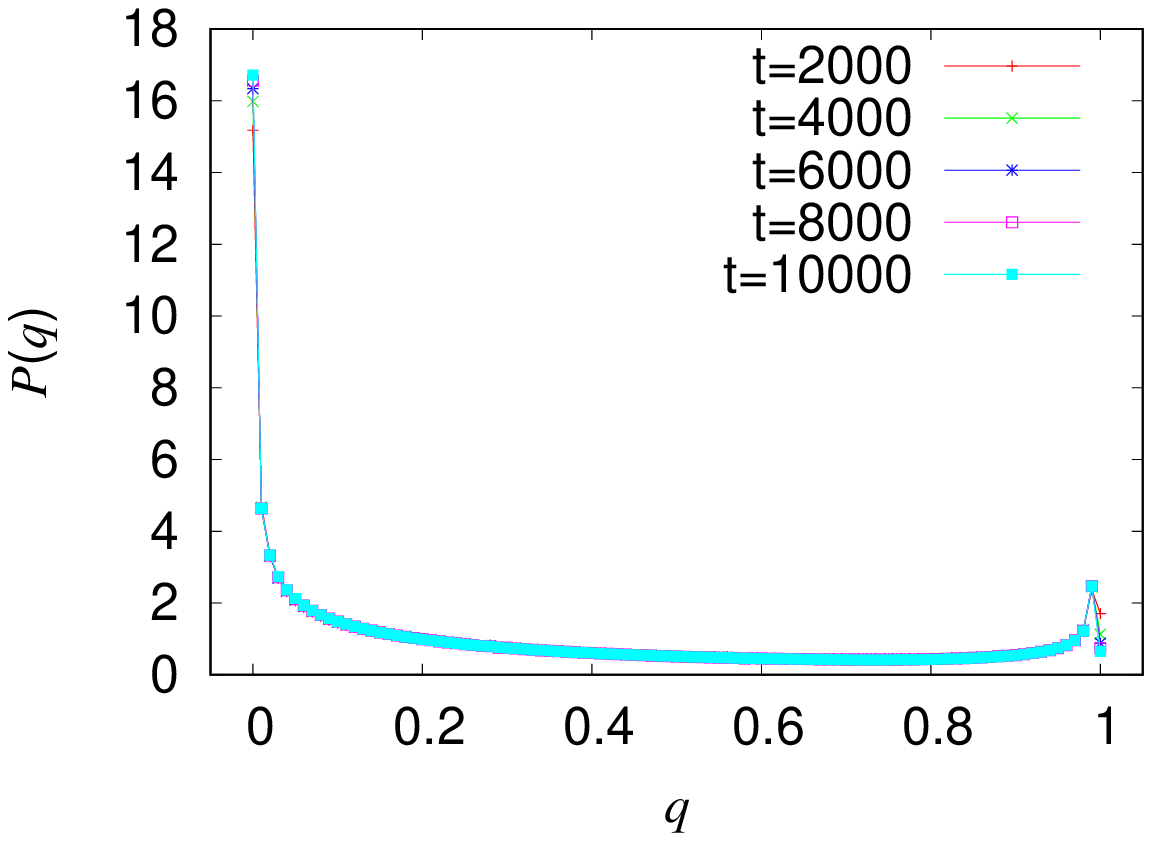}
\includegraphics[clip, width=8.0cm]{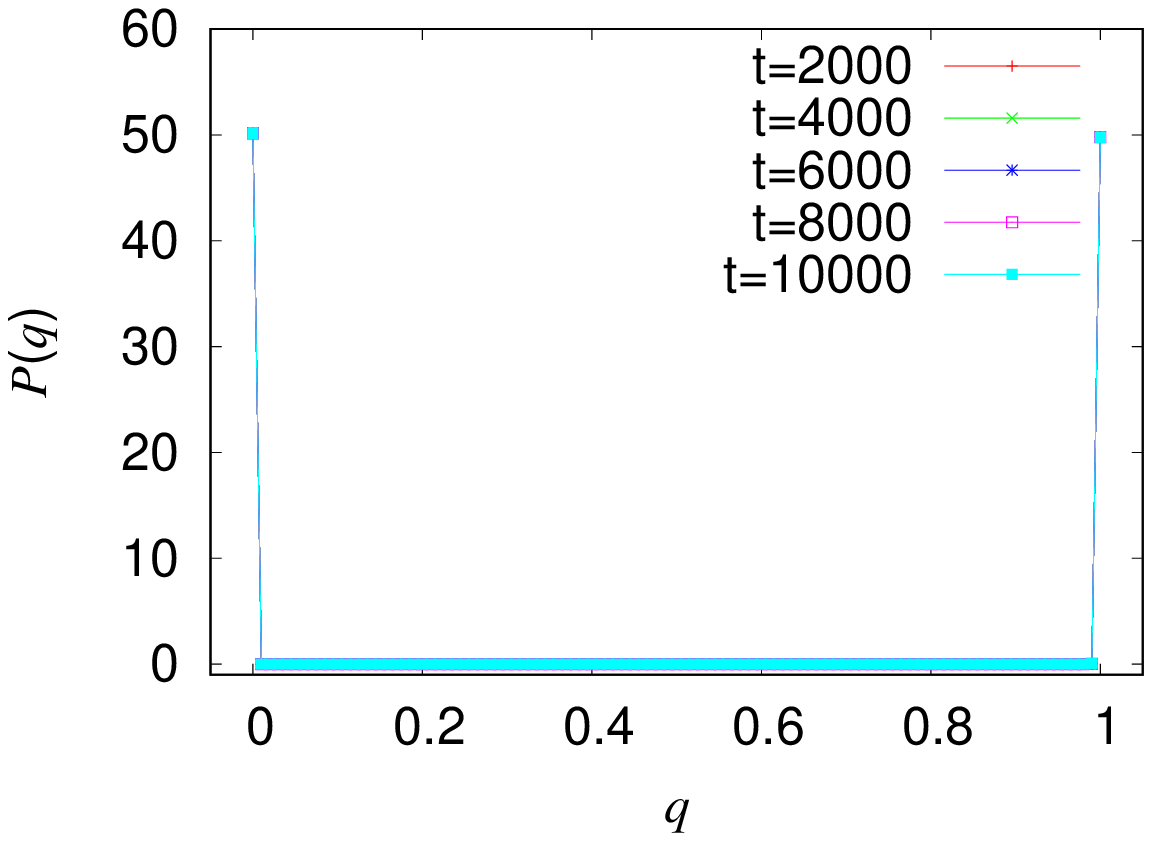}
\caption{The overlap distribution $P(q)$ at $\beta=2$ under the initial distributions $P_0(j)=\delta_{j,1}$ (left) and $P_\mathrm{eq}(j)$ (right).}
\label{fig:ovhisttrap_0.5}
\end{figure}

Before ending this paper, we make two remarks.
First, RSB in trajectories can be confirmed more directly by the distribution of overlap
\begin{eqnarray}
 P(q) &=& \mathbb{E} \left[ \left\langle \delta\left( q - \frac{1}{\tau} \sum_{l=1}^\tau \delta_{j_l^{(1)}, j_l^{(2)}} \right) \right\rangle \right].
\end{eqnarray}
In spin-glass theory, RSB is described by the existence of non-trivial peaks in $P(q)$ in addition to the peak at $q=0$ \cite{MPV1987}.
Similarly, in our problem, when two systems freeze into the same trajectory, overlap takes a finite value, while overlap takes zero when two systems freeze into different trajectories respectively.
We display the distributions of overlap $P(q)$ for temperatures $\beta=0.5$ and $\beta=2$ in Fig. \ref{fig:ovhisttrap_2.0} and Fig. \ref{fig:ovhisttrap_0.5}, respectively, while the results for different temperatures are qualitatively similar in each temperature regime.
Results for two initial distributions $P_0(j)=\delta_{j,1}$ and $P_0(j)=P_\mathrm{eq}(j)$ are displayed.
Numerical calculation is performed for space size $M=4000$, the number of realizations of a trajectory $N_\mathrm{p}=80000$, and the number of realizations of a random potential $N_\mathrm{d}=1000$.
It should be noted that we focus on the time region $\tau \ll \tau_\mathrm{eq}(M)$ with (\ref{eq:eqtime_low}).
We observe that a non-trivial peak appears in $P(q)$ in the low temperature phase, although there is only one trivial peak in the high temperature phase.
%It should be noted that many data points are located on the vertical and horizontal axes, and data for different times are also overlapped in these figures.
The coexistence of two peaks in $\beta>1$ implies RSB in trajectories for both initial conditions.
We note that the form of $P(q)$ depends on the initial distribution because the system never relaxes to the equilibrium state in the large $M$ limit, as in the results for $\left\langle q \right\rangle_\epsilon$.
These results provide further evidence for our main claim that RSB occurs for the case where the limit $\tau \rightarrow \infty$ is considered after $M \rightarrow \infty$ is taken.

The second remark is made on the universality of RSB in trajectory space.
Our result suggests the possibility that other subdiffusive systems also exhibit RSB in trajectory space.
One of the candidates is a particle in a random potential with logarithmic correlations.
In statics, this model exhibits localization into a few states, and relation to RSB was suggested \cite{CarLeD2001}.
In dynamics, the dynamical transition between two subdiffusive phases was observed, and a non-equilibrium splitting of the thermal distribution of the diffusing particle into a few packets was expected in the low temperature phase \cite{CasLeD2001}.
This model will be studied in terms of RSB in trajectory space in future.

In this paper, for the trap model, we studied the localization phenomenon in the low temperature phase $\beta>1$ in terms of statistical mechanics of trajectories.
We have numerically found that RSB in trajectories occurs regardless of the order of the two limits $M \rightarrow \infty$ and $\tau \rightarrow \infty$ in calculation of the cumulant generating function of overlap, while there is no localization in trajectory space in the high temperature phase $\beta<1$.
Although equilibrium statistical mechanics of the trap model is singular, this model exhibits the second-order phase transition in the framework of statistical mechanics of trajectories.

% acknowledgements
\ack
The authors thank M Iwata, A Dechant, A Baule, M Itami and T Haga for valuable discussions.
The present study was supported by KAKENHI (Nos. 25103002 and 26610115) and a Grant-in-Aid for JSPS Fellows (Nos. 14J00081 and 16J00178).


\section*{References}
\begin{thebibliography}{99}
 % subdiffusion in nature
 \bibitem{TMTet2004} Toli\'{c}-N{\o}rrelykke I M, Munteanu E L, Thon G, Oddershede L and Berg-S{\o}rensen K, 2004 {\it Phys. Rev. Lett.} {\bf 93} 078102
 \bibitem{WGRet2004} Wong I Y, Gardel M L, Reichman D R, Weeks E R, Valentine M T, Bausch A R and Weitz D A, 2004 {\it Phys. Rev. Lett.} {\bf 92} 178101
 \bibitem{GolCox2006} Golding I and Cox E C, 2006 {\it Phys. Rev. Lett.} {\bf 96} 098102
 \bibitem{SzyWei2009} Szymanski J and Weiss M, 2009 {\it Phys. Rev. Lett.} {\bf 103} 038102
 \bibitem{BIKet2009} Bronstein I, Israel Y, Kepten E, Mai S, Shav-Tal Y, Barkai E and Garini Y, 2009 {\it Phys. Rev. Lett.} {\bf 103} 018102
 \bibitem{SenMar2010} Senning E N and Marcus A H, 2010 {\it Proc. Natl. Acad. Sci.} {\bf 107} 721
 \bibitem{WSTK2011} Weigel A V, Simon B, Tamkun M M and Krapf D, 2011 {\it Proc. Natl. Acad. Sci.} {\bf 108} 6438
 \bibitem{JTBet2011} Jeon J H, Tejedor V, Burov S, Barkai E, Selhuber-Unkel C, Berg-Sorensen K, Oddershede L and Metzler R, 2011 {\it Phys. Rev. Lett.} {\bf 106} 048103
 \bibitem{PSCet2014} Parry B R, Surovtsev I V, Cabeen M T, O'Hern C S, Dufresne E R and Jacobs-Wagner C, 2014 {\it Cell} {\bf 156} 183
 
 % subdiffusion model
 \bibitem{BouGeo1990} Bouchaud J P and Georges A, 1990 {\it Phys. Rep.} {\bf 195} 127
 \bibitem{MetKla2000} Metzler R and Klafter J, 2000 {\it Phys. Rep.} {\bf 339} 1
 \bibitem{BJMB2011} Burov S, Jeon J H, Metzler R and Barkai E, 2011 {\it Phys. Chem. Chem. Phys.} {\bf 13} 1800
 \bibitem{HofFra2013} H\"{o}fling F and Franosch T, 2013 {\it Rep. Prog. Phys.} {\bf 76} 046602
 
 % weak ergodicity breaking
 \bibitem{HBMB2008} He Y, Burov S, Metzler R and Barkai E, 2008 {\it Phys. Rev. Lett.} {\bf 101} 058101
 
 % trap model
 \bibitem{Bou1992} Bouchaud J P, 1992 {\it J. Phys. France} {\bf 2} 1705
 \bibitem{MonBou1996} Monthus C and Bouchaud J P, 1996 {\it J. Phys. A Math. Gen.} {\bf 29} 3847
 \bibitem{BerBou2003} Bertin E M and Bouchaud J P, 2003 {\it Phys. Rev. E} {\bf 67} 026128
 
 % RSB in trajectories
 \bibitem{UedSas2015} Ueda M and Sasa S, 2015 {\it Phys. Rev. Lett.} {\bf 115} 080605
 
 % RSB
 \bibitem{MPV1987} M\'{e}zard M, Parisi G and Virasoro M A, 1987 {\it Spin glass theory and beyond} (Singapore: World Scientific)
 
 % statistical mechanics of trajectories
 \bibitem{BecSch1993} Beck C and Sch\"{o}gl F, 1993 {\it Thermodynamics of chaotic systems: an introduction} (Cambridge: Cambridge University Press)
 \bibitem{GJLPDW2007} Garrahan J P, Jack R L, Lecomte V, Pitard E, van Duijvendijk K and van Wijland F, 2007 {\it Phys. Rev. Lett.} {\bf 98} 195702
 \bibitem{HJGC2009} Hedges L O, Jack R L, Garrahan J P and Chandler D, 2009 {\it Science} {\bf 323} 1309
 \bibitem{JacGar2010} Jack R L and Garrahan J P, 2010 {\it Phys. Rev. E} {\bf 81} 011111
 \bibitem{GKLT2011} Giardina C, Kurchan J, Lecomte V and Tailleur J, 2011 {\it J. Stat. Phys.} {\bf 145} 787
 \bibitem{LLKT2013} Laffargue T, Lam K D N T, Kurchan J and Tailleur J, 2013 {\it J. Phys. A: Math. Theor.} {\bf 46} 254002
 \bibitem{LSTW2015} Laffargue T, Sollich P, Tailleur J and van Wijland F, 2015 {\it Europhys. Lett.} {\bf 110} 10006
 
 % REM
 \bibitem{BouMez1997} Bouchaud J P and M\'{e}zard M, 1997 {\it J. Phys. A: Math. Gen.} {\bf 30} 7997
 
 % KS entropy
 \bibitem{LAW2005} Lecomte V, Appert-Rolland C and van Wijland F, 2005 {\it Phys. Rev. Lett.} {\bf 95} 010601
 \bibitem{LAW2007} Lecomte V, Appert-Rolland C and van Wijland F, 2007 {\it J. Stat. Phys.} {\bf 127} 51
 
 % KS entropy of trap model
 \bibitem{IwaSasup} Iwata M and Sasa S, 2010 {\it Meeting abstracts of the Physical Society of Japan} {\bf 65}(1-2) 378 (in Japanese)
 
 % equilibrium of trap model
 \bibitem{Der1997} Derrida B, 1997 {\it Physica D} {\bf 107} 186
 
 % RSB in directed polymer
 \bibitem{Mez1990} M\'{e}zard M, 1990 {\it J. Phys. France} {\bf 51} 1831
 
 % directed polymer
 \bibitem{HalZha1995} Halpin-Healy T and Zhang Y C, 1995 {\it Phys. Rep.} {\bf 254} 215
 
 % infinite density
 %\bibitem{SchBar2015} Schulz J H P and Barkai E, 2015 {\it Phys. Rev. E} {\bf 91} 062129
 
 % potential with logarithmic correlation
 \bibitem{CarLeD2001} Carpentier D and Le Doussal P, 2001 {\it Phys. Rev. E} {\bf 63} 026110
 \bibitem{CasLeD2001} Castillo H E and Le Doussal P, 2001 {\it Phys. Rev. Lett.} {\bf 86} 4859
 
 % statistical mechanics
 %\bibitem{Tassm} Tasaki H, 2008 {\it Statistical mechanics} (Tokyo: Baifukan) (in Japanese)
 
\end{thebibliography}
\end{document}